# The *Beyond the Fence* Musical and *Computer Says Show* Documentary

**Simon Colton**[1,2], **Maria Teresa Llano**[1], **Rose Hepworth**[1], **John Charnley**[1],
**Catherine V. Gale**[3], **Archie Baron**[3], **François Pachet**[4], **Pierre Roy**[4], **Pablo Gervás**[5],
**Nick Collins**[6], **Bob Sturm**[7], **Tillman Weyde**[8], **Daniel Wolff**[8] and **James Robert Lloyd**[9]

[1]Goldsmiths College, London, UK    [2]Falmouth University, UK    [3]Wingspan Productions, London, UK
[4]Sony Computer Science Laboratory, Paris, France    [5]Universidad Complutense, Madrid, Spain    [6]Durham University, UK
[7]Queen Mary University of London, UK    [8]City University London, UK    [9]Qlearsite Organisational Science, UK

**Abstract**

During 2015 and early 2016, the cultural application of Computational Creativity research and practice took a big leap forward, with a project where multiple computational systems were used to provide advice and material for a new musical theatre production. Billed as the world's first 'computer musical ... conceived by computer and substantially crafted by computer', *Beyond The Fence* was staged in the Arts Theatre in London's West End during February and March of 2016. Various computational approaches to analytical and generative sub-projects were used to bring about the musical, and these efforts were recorded in two 1-hour documentary films made by Wingspan Productions, which were aired on Sky Arts under the title *Computer Says Show*. We provide details here of the project conception and execution, including details of the systems which took on some of the creative responsibility in writing the musical, and the contributions they made. We also provide details of the impact of the project, including a perspective from the two (human) writers with overall control of the creative aspects the musical.

## Introduction

There are very few types of cultural creations more complex than a musical theatre show. The very basics required to design and produce a new musical include the creation of: an overall concept, a narrative arc, characters, dialogue and plot lines; music, orchestration and lyrics; set design, artwork, and lighting routines; and finally crafting the overall performance through rehearsals and direction, along with producing advertising material to promote the event. While stand-alone computational systems able to create artefacts in many of these areas exist, it is quite far beyond the state of the art of Computational Creativity practice to imagine the full creation of a musical via a single creative system or even by multiple systems under some automated directorial control. Hence a fully computer generated musical stands as a grand challenge for our field, which can serve to drive technologies forward, broaden public understanding of Computational Creativity and highlight new areas of research.

In late 2014, Wingspan Productions, a London-based television production company (wingspanproductions.co.uk), began to develop the idea of devising and filming an experiment in which software was used as much as possible to take on some of the creative responsibilities of writing a musical theatre piece. The project was carried through to completion successfully, and the end results were: (a) the staging of the *Beyond The Fence* musical in the Arts Theatre in London's West End in February/March 2016, the design of which was heavily influenced by software analysis of musical theatre data, and by creative software which contributed an original concept, plot lines, lyrics and music, and (b) two 1-hour television documentaries for Sky Arts entitled *Computer Says Show*, charting the making of the musical, with emphasis on how software was used in the creative process and on the journey of discovery that the two (human) writers went on within an experiment where they were asked to use the software's advice and output as much as (humanly) possible.

Below, we provide first person overviews of the conception of the project, from the Wingspan Productions team. Following this, we describe the analytical and generative systems which contributed material used to both guide the writing of the musical and explicitly in the final production, in terms of how the systems operate, how they were employed and what they produced. We also provide details of the cultural impact of the staging of the show and the airing of the two documentaries. This includes details of the press coverage, critical reviews and a perspective on working with creative software by the two writers of the show. We conclude by highlighting the importance of projects such as this for the field of Computational Creativity.

## Project Conception

The driving forces behind the entire project were Archie Baron, Executive Producer, and Catherine Gale, Series Producer/Director of *Computer Says Show* and *Beyond the Fence*, from Wingspan Productions. The following are first person accounts of the project conception from the perspectives of this production team:

**The Idea** (Archie Baron)
The musical *Beyond The Fence* is the unlikely offspring of two television documentaries we made in 2014. *The Joy of Logic* (BBC) explored the hypothesis – based on Alan Turing's notion that the brain and a computer are both essentially information processing systems governed by logical rules – that one day we could in theory program a computer to reproduce and rival all of human thought. *Our Gay Wedding – The Musical* (Channel 4) was just that: Benjamin Till and Nathan Taylor's wedding on the first day same-sex weddings were legalised, composed by the grooms and broadcast as a joyous sung-through musical. An improbable mash-up of these two projects led us, in discussion with the British television channel Sky Arts, who were interested in what might underpin success and failure in the arts, to the basic idea for the two-part documentary series *Computer Says Show*: could we use machine learning to analyse what makes a hit musical, then team up with computers to create and stage one?

From the outset, the project was conceived as both a serious experiment and a provocative and audacious event. It was also geared as much towards documenting the process as investing in the outcome. This 'experiment' and the resulting documentaries and musical were above all designed to provoke a debate. What constitutes success in creating music or stories? What parts of the creative process can be automated? How do scientists, writers, composers and audiences respond to work created in collaboration with computers? What might the general public make of the leading edge of this field – in terms of the science, technology and philosophy that underpins it? This is a debate worth having. What was it Alexander Graham Bell said? "I truly believe that one day there will be a telephone in every town in America." The future – as with so much technology – is probably nearer and more extraordinary than most of us imagine.

**The Practicalities** (Catherine Gale)
Bringing together the team to create Beyond The Fence was one of the most fascinating and fulfilling challenges of my career so far – in science or programme making. For brevity, it can be broken down into three main stages:

1. Initial research: scoping out the leading edge in the theory and practice of computer generated art – in particular music, story, dialogue/lyrics.
2. Consortium formation and pre-production: bringing together a group of academics interested in turning their attention to musical theatre, to contribute to the project such that both the musical itself and their own research projects would benefit. Significant data gathering and annotation was required, as well as collaborative discussions about methodology and goals, and documentary filming to tell the story of the 'experiment' as it unfolded (see figure 1).
3. Writing the show: all the data analysed (Cambridge dataset), systems developed/made available (WHIM for musicals, ProperWryter, The Cloud Lyricist, Flow Composer) or material generated (from Nick Collins) had to be used by the writers, to create the new show. These processes were also documented on film. The writers worked to a set of guidelines that ensured that computer-generated content would remain the core 'raw material' for the show. This enabled a story to be told around the experience of the (human) writers having to work with computer-generated material when putting the show together; the actors and creative team's experience of working with it; and the audience's response to it. It also meant that we could explore in detail the science behind each process, whether predictive analytics, machine listening, algorithmic composition, plot analysis and generation, or lyric generation.

An exciting feature of this project is that, having reached the finish line in terms of the musical and the documentaries, we now have a 'case study' in hand, which perhaps only answers a small number of questions in and of itself, but poses far more about the role that computational systems can, will and should play in the future of all of our lives – as creators and consumers of both art and technology.

**The Guidelines** (Archie Baron)
One of the toughest challenges for the project was framing sufficiently rigid protocols that articulated for everyone involved the basic framework for the project, while having sufficient flexibility that we could document and observe an unfolding process about whose outcomes we were extremely uncertain until very late in the day. We codified these protocols into a formal set of ground rules which we asked the writers and everyone involved in the creative team to work to in an attempt to ensure that computer-generated content would remain the core 'raw material' for the new show. The fact that the project had two key aims which were not necessarily compatible – to model a hit musical using the Cambridge data (see below) which West End theatregoers would buy tickets for and to maximise the computer-generated content therein – became an editorial and creative challenge (and an important narrative angle for the documentaries).

## Contributing Systems and Application Details

As portrayed in figure 1, how the systems used in the experiment contributed to the musical is somewhat complicated. In overview, two writers were in overall creative control of the writing of the musical: Benjamin Till and Nathan Taylor. Software contributed to the music, lyrics and story of the musical, and also provided overall guidance to the writers, which was referred to throughout the process.

A corpus analysis was used to identify factors associated with successful and unsuccessful shows, which influenced all other aspects of the writing process. The story for the musical was then developed from an original concept produced by *The WhatIf Machine* ideation engine developed at Goldsmiths College, and output from *ProperWryter*, a plot generation system developed at Universidad Complutense, both of which are described below. Data input to ProperWryter included an annotated dataset of stories, created specifically for the project. The writing of the musical score was informed by a music information retrieval (MIR) analysis of the musicological features of successful and less successful musicals, produced via a study at City University London and Queen Mary University of London, described below.

The analysis informed the first of two algorithmic composition techniques, a corpus based generative system developed at Durham University, described below. A second generative system called *Flow Composer*, from the Sony Computer Science Laboratory, was also used, producing new music in an interactive way, via style modelling, as described below. Finally, also described below, new software called *The Cloud Lyricist* was developed which used a neural network approach trained on musical theatre lyrics to generate text segments which were carefully selected by the writers. The music, lyrics and story inputs from the computational systems were combined by Till and Taylor into the final musical, guided by the machine learned analysis, then shared with the director, Luke Sheppard, and team, for subsequent development, rehearsals and staging.

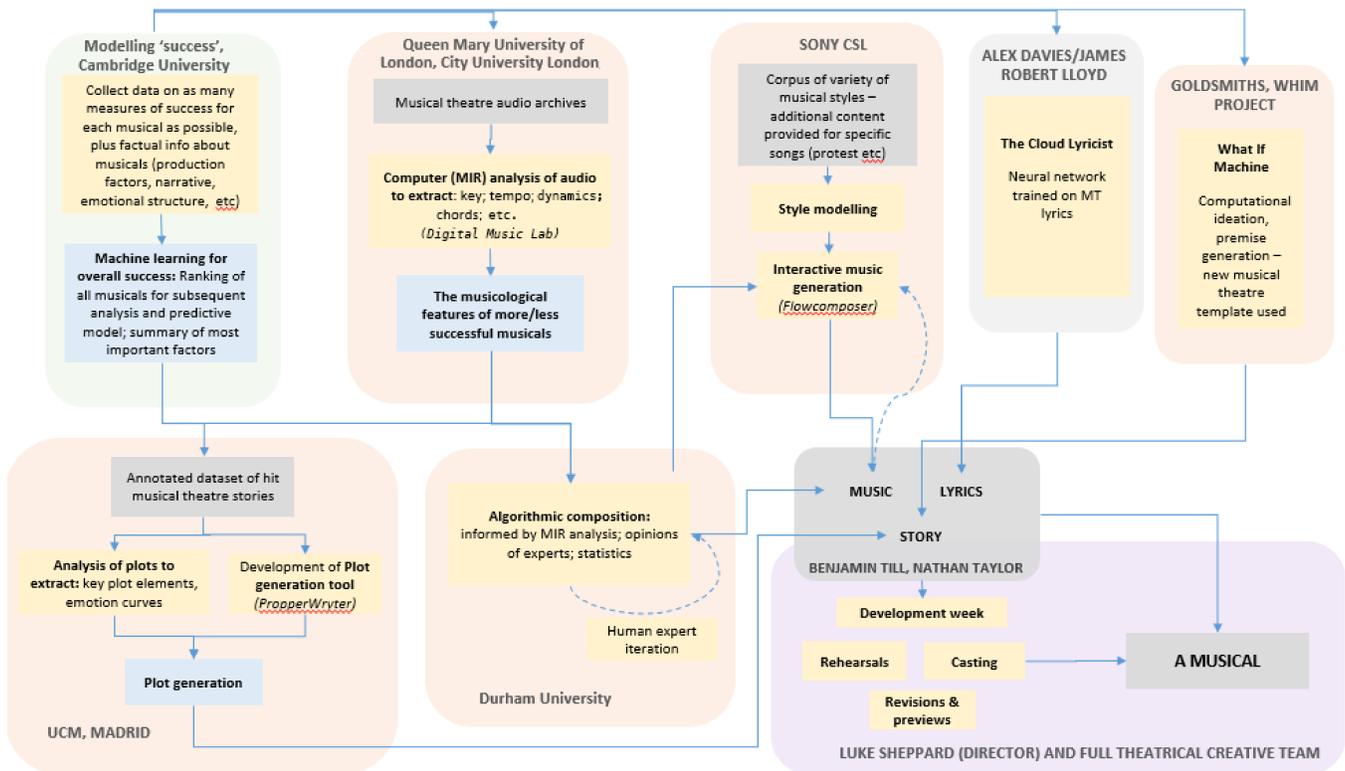

Figure 1: Research teams and their contributions to the project.

## The Statistics for Success in Musical Theatre

The first stage was the study 'Musical Theatre by Numbers' combining descriptive statistics, machine learning and predictive analytics conducted at Cambridge University by James Robert Lloyd, Alex Davies and David Spiegelhalter. They built a corpus of information about 1696 musicals, including 946 synopses and in-depth surveys annotating individual shows. All musicals were measured for both commercial and critical success based on length of run and awards won, and then classified into one of 4 categories (hits: commercial and critical success; crowd pleasers: commercial success only; critically acclaimed: awards but shorter runs; flops: neither commercial nor critical success). The aim was to discover the factors that are associated with (and might be predictive of) success, via exploratory data analysis using logistic regression and decision trees.

The relative importance of high level features such as cast size, the gender of the lead, musical styles, geographic and temporal settings, star power, a happy ending, the incidence of comedy, death, spectacle, dance within a show, were all investigated. A thematic study of shows was also achieved through synopsis text analysis. The emotional journey (or story arc) throughout a selection of 52 representative shows was also annotated and analysed. Volunteers listened to soundtrack recordings of complete musicals they knew well, recording how they felt about each song in ten different emotional classes (to attempt to capture the emotional trajectory of musicals in terms of energy and vitality; love and tenderness; tension and sadness; comedy). Clear differences survived the averaging of these emotional arcs to distinguish hits from flops, and these became key features defining the target emotional structure of what became *Beyond the Fence*. The results of this analysis were presented to the writers as a presentation (with an example graphic from the presentation given in figure 2) focusing on the key decisions that might increase the chances of writing a hit.

## The WhatIf Machine

In the creative arts and industries, the production of fictional ideas around which to write stories, paint pictures or design advertisements, is an essential activity. Motivated by this, in the European WHIM project (www.whim-project.eu), via the building of The WhatIf Machine, we are undertaking the first large-scale study of how software can invent, evaluate and express fictional ideas of real cultural value.

We refer to fictional ideas as modifications of knowledge where the perceptions we hold about existing concepts of the world are altered and new representations are produced. The Goldsmiths approach in the WHIM project largely consists of applying controlled alterations and combinations of facts from a knowledge base (KB), represented as triples which relate two concepts. We have developed a set of ideation methods which have been reported in (Llano et al. 2016). For instance, given the natural language template: $t$ = *What if a X learned how to Y?*, the simple ideation method:

$$im(t) = \{r \mid \langle c_1, NotCapableOf, c_2 \rangle \in KB \quad (1)$$
$$\land \ r = instantiate(t, [X = c_1, Y = c_2]\}$$

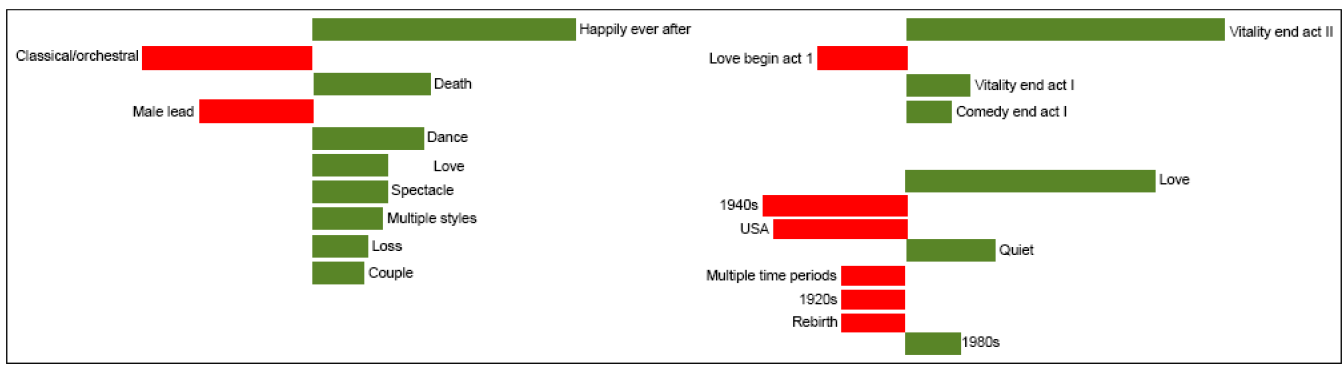

Figure 2: Statistical analysis of (green) hits and (red) flops, with bar length indicating the propensity of the factor in the respective category.

produces a fictional idea using $t$, for facts in the KB that focus on the *NotCapableOf* relation. For instance, given the fact ⟨*dog,NotCapableOf, ride_horse*⟩, (1) yields the fictional idea: *What if a dog learned how to ride a horse?* A set of conditions and heuristics are applied to narrow down the input knowledge and to perform more complex combinations of facts. Moreover, the generated ideas are evaluated against a set of narrative measures, which are then compared with an audience model that returns the list ranked according to a machine learned predictor of value (details omitted).

The Cambridge thematic analysis identified four general themes which were passed on to the WHIM team: *journey, aspiration, love* and *lost king*, and for each theme a set of keywords were also identified. Additionally, the writers summarised a set of musical synopses, from hits, crowd pleasers, critically acclaimed and flops, into what-if style sentences, which acted as targets. Finally, we were provided with a description of Bookers' seven basic plots (Booker 2004) as standard plot types followed in musical theatre. Because of the short time span to complete the experiment, we focused on three of these plot types, namely: *rags to riches, quest* and *rebirth*. Note that, while – as per figure 2 – rebirth was identified as a negative factor, it was included due to handover timing issues. We combined the plot types with the themes mentioned above in order to produce different types of fictional ideas in the musical context. All this data was used to: (i) retrieve further input knowledge from different linguistic resources (using the theme keywords), (ii) create template what-if premises (using the musical premises), and (iii) select the components required by the premises to be generated by the system (using the plot types).

To illustrate, following the description of the *Rebirth* plot type, which states: *'The protagonist is a villain or otherwise unlikable character who redeems him/herself over the course of the story'*, and the theme *love*, we knew that the character of the statement should carry out a transformative action in order to gain someone's affections. The transformative action was selected as a combination of data obtained from the stereotypical properties of selected characters and the further knowledge obtained through the keywords. Figure 3 shows a screenshot of the musicals area of the web interface to The WhatIf Machine, which shows some sample Rebirth/Love outputs.

The Wingspan Productions team used The WhatIf Machine to choose and print out 600 ideas suitable as the overall concept for the musical. The choosing of candidates from this set was filmed on a stage at Goldsmiths, where all 600 were laid out. Some ideas considered were not practical, such as *"What if there was a poor boy who was born with a horn, which made him good at communicating, so he went on to become a famous slave"*. However, the writers were able to take away a shortlist of ideas for further consideration. Guided by the Cambridge analysis that the musical should be set in a conflict of the 1980s, not in America, and involve a female protagonist, they finally settled on the idea: *"What if there was a wounded soldier who had to learn to understand a child in order to find true love?"* Elements from this phrase, plus other constraints (e.g. death, loss) were typed into a search engine and one of the top hits was a songbook from the Greenham Common Womens' anti-nuclear peace camp (www.yourgreenham.co.uk). Ultimately, this led to the writers' interpretation of the idea, whereby a male soldier who has been wounded and posted to Greenham Common nuclear base in the UK befriends the mute child of a female protester, which ultimately leads to him and the protester falling in love.

### PropperWryter

PropperWryter is a program that generates the narrative structure for a single plot line, described in terms of a vocabulary of abstract representations of events that may happen in such a plot. It evolved from the Propper system (Gervás 2015), which generated plot structures for Russian folk tales based on Propp's (1968) *Morphology of the Folktale*.

Propp identified a set of regularities across a corpus of Russian folk tales in terms of *character functions*, understood as acts of the character, defined from the point of view of their significance for the course of the action. To extend this approach to the generation of plots for musicals, the vocabulary was adapted and extended to cover the range of acts of characters typically involved in musicals, and data had to be collected on how elements from this customised vocabulary appeared in existing musicals, and on how these elements interacted with one another. The new vocabulary for describing musical theatre plot was built as a new specific set of abstractions in the spirit of character functions – renamed as *plot elements* to avoid confusion – and constructed from a number of sources: the original set of Propp's character functions; additional abstractions mined from alternative sources in the literature on narrative (Gervás, León, and Méndez 2015); and abstractions specific to the description of the plot of musical theatre. Knowledge resources on how this vocabulary is used in existing musicals were compiled from the results of an annotation effort of musical plots carried out by 35 volunteers at the University of Surrey, supervised by Julian Woolford, Programme Leader for the MA in Musical Theatre, and the Wingspan Productions team.

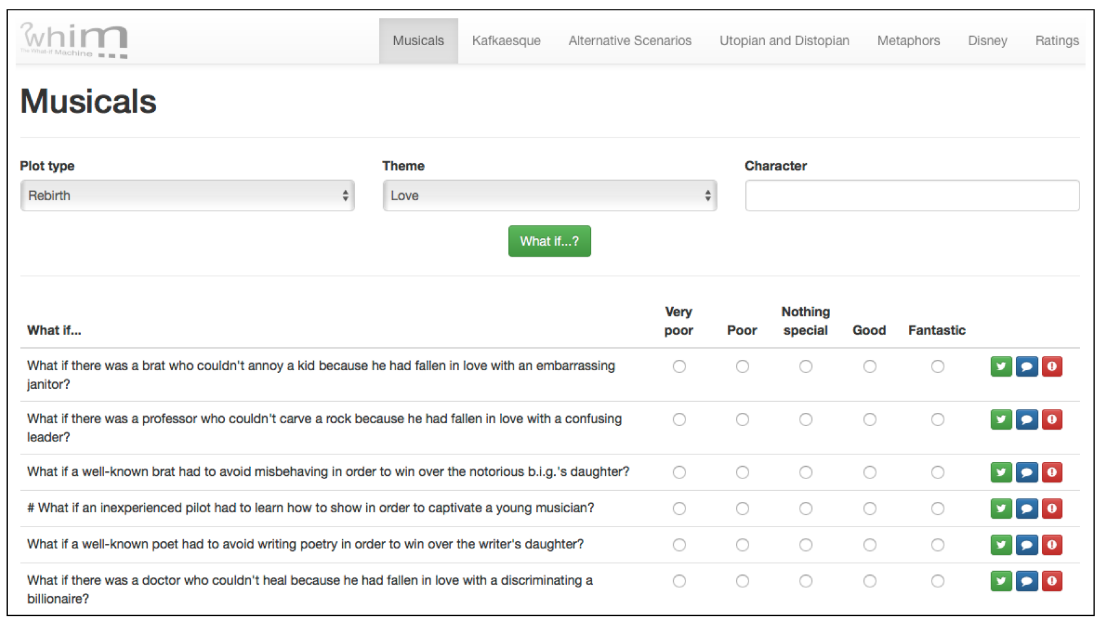

Figure 3: The web interface for The WhatIf Machine.

The PropperWryter system relied on these knowledge resources to construct sequences of plot elements – from the new vocabulary – that describe a plot line. On reviewing earlier versions of the prototype, which allowed a user to provide an initial brief to drive the plot construction process, the writers requested that this functionality be disabled, to avoid any risk of any features in the output being attributed to the input rather than the machine's efforts. During the writing phase, the writers used PropperWryter to generate a number of possible plots and selected one that started with an Aspiration plot element. The plot structure of *Beyond the Fence* is, therefore, entirely underpinned by a ten-point arc generated by the software.

**Music Analysis with the Digital Music Lab**

To study the acoustic and musical characteristics of musicals from the four categories established by the Cambridge team, we (Bob L. Sturm, Tillman Weyde and Daniel Wolff) employed the Digital Music Lab (DML) infrastructure – the outcome of a recent project developed for large-scale music analysis in a collaboration of City University London, Queen Mary University of London, UCL and the British Library (Weyde et al. 2014). The musicals project encapsulates an excellent use case from the DML: we needed to study a massive amount of music material from a variety of different perspectives in a short amount of time. We extracted low-, mid- and high-level features from our corpus of 77 full-length commercial recordings of musicals – over 130 hours of sound recording of diverse styles and from different historic periods – and analysed dimensions such as loudness, brightness, tempo, dynamics, key, and harmonies.

Our analysis delivered expected results: we find a predominant use of major keys across all categories, and we find a lack of strong prediction power of these features (which for the most part are far from describing the experience of music) for the four categories. Averaging the features across musicals in each of two categories (flop (N=23) or non-flop (N=54)) – taking care to normalise the time scale of the musicals – showed some intriguing patterns. Figure 4 shows the change in dynamics (loudness) from the mean of a track – essentially (de)crescendi in a track. We see a tendency in the final 25% music of flops for tracks to become increasingly more dramatic. We also see that the final song in flops tend to be faster than average. At a higher level, we found that flops in our corpus tend to have more harmonic progressions containing the subdominant.

Our results must be handled with caution for at least two reasons: 1) we do not demonstrate any causation, e.g., that the use of the subdominant affects the success of a musical; and 2) the flops in our corpus are the "best kind", i.e., successful enough to be commercially released as a recording. The computational analysis of music remains far away from human listening experience and expertise. Nonetheless, our involvement in this project shows the success of the DML infrastructure for studying the questions posed by the production team, as well as providing high-level musical features for the modeling and generation of music.

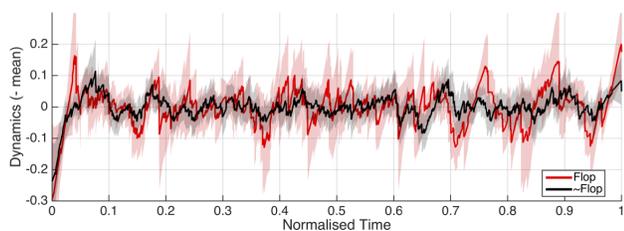

Figure 4: The tempo relative to track average over normalised duration for flop and non-flops classes.

## Claude-Machine Schönbot

The music generating program Claude-Machine Schönbot (formerly known as 'Android Lloyd Webber') was used to produce the majority of the computer generated music for Beyond the Fence. It is based on a corpus of chord changes from hit musicals, and an additional corpus of transcribed leadsheets (main melody and accompanying chords) from hit musicals. Which musicals were denoted hits, and the chord changes extracted from cast recording audio files (53 musicals, 1124 audio files, around 53 hours of audio), arose from other groups involved in the project (see above). Markovian machine learning (of variable order, via prediction by partial match (Pearce and Wiggins 2004) is deployed, alongside hard-coded rules informed by the statistics of the melodic examples. As all corpus-based algorithmic composition makes representational assumptions on the nature of music data analysed and generated, and given the pragmatics of theatre shows with hard deadlines, the involvement of explicit rules – which were developed via feedback from the writers – was seen as acceptable. The program could generate melody and chord leadsheets as pure music, or with rhythmic phrases constrained by lyrics provided, following a basic metrical stress analysis of scansion.

## Flow Composer

Flow Composer is an online lead sheet editor with a composition module relying on style-imitation: the system builds a *statistical model* from a corpus of lead sheets and generates new musical material in the corpus style (Pachet and Roy 2014). Composing a lead sheet is achieved in a progressive way, involving several rounds of user-system interactions. The user starts with an empty or partial lead sheet (see figure 5). First, Flow Composer is repeatedly used to generate music for the selected zone until something interesting comes up. The user deselects the part they want to keep and uses the generator on the remaining selected zone. At all times, the user may freely modify the current lead sheet and use the history to go back to a previous state. The music can be listened to using an audio engine producing natural renderings in many musical styles. Flow Composer samples random sequences subject to constraints from the statistical model (Papadopoulos et al. 2015). The musical elements to generate (selected zone) are treated as random variables and constraints represent the notes and chords that the user wants to keep unchanged (non-selected zones).

Flow Composer was initially designed for jazz, Brazilian, and pop music generation. It is coupled with a large database containing 13,000 songs in these styles (Pachet, Suzda, and Martinez 2013). The writers thought it was appropriate to train the system with songs composed by the women of Greenham Common (a songbook is available at yourgreenham.co.uk). They chose six songs that were added to the database. With such a small corpus (75 bars in total), the system often fails to generate music fitting the user's input, leading to frustrating interactions. We changed the training mechanism so that the system learns the user's input, i.e., as the user enters new music, or modifies the system's output, the corpus is enriched and the model is retrained, reducing failures. We also added the ability to add transposed copies of original songs in the corpus, to create more diversity.

The writers expressed the need to control some musical parameters of the generated sequences. The sampling mechanism guaranties that the generated music has the same statistical properties as the training corpus, but provides no way to control the generation. We added controls for five musical parameters: harmonic tightness, average melodic interval, average note duration, proportion of rests, and frequency of chord changes. Each parameter is controlled using a slider widget on the graphical user interface (see top of figure 5). These criteria are taken into account by a simple generate-and-test mechanism: several sequences are sampled, and the one that best fits the criteria is returned. The speed of the sampling mechanism makes it possible to generate thousands of sequences in a few seconds.

The writers used Flow Composer for three songs: *Scratch that Itch*, *We Are Greenham* and *Unbreakable*, modifying the raw output, for instance to fit the rhythm to the lyrics or to create an ending to the melody, which conforms to the type of iterative interaction Flow Composer was designed for. This also suggests interesting directions to improve the system in the future. Flow Composer is currently being used for the composition of a pop music album.

## Clarissa, The Cloud Lyricist

Once the data and software-guided basic structure, setting, key features, central idea and plot arc were in place, the writers proceeded to prepare the spoken dialogue for the show. Alex Davies and James Robert Lloyd, who had already had some success with computer generated poetry, trained a recurrent neural network language model (based on Andrej Karpathy's Char-RNN: karpathy.github.io/2015/05/21/rnn-effectiveness/) on a corpus of some 7,000 musical theatre lyrics (after first pre-training on a portion of Wikipedia and a corpus of 10,000 poems) and built The Cloud Lyricist (or Clarissa as 'she' became known in the rehearsal room). Clarissa supplied an initial 1,000,000 characters of lyrics for the writers and computer lyric dramaturge Kat Mace to draw from and incorporate into the songs. They later developed the system with a web interface with a 'creativity factor' setting which could be set by the user to increase the abstraction level (from 0 to 1) in the language produced (using the temperature parameter in Char-RNN). The setting of 0.6 or 0.7 proved to be the most useful.

The lyric writing process required time-consuming trawls through Clarissa's limitless oeuvre, usually to find brief (usually part line or single line) stretches of usable lyrics which could be incorporated with other computer lines and/or lines from the writers. In total, almost a quarter of the lyrics in *Beyond the Fence* were computer generated (ranging across the 16 numbers from 6% in the song *Graceful* to 32% in *So Much to Say*). Close textual analysis of the libretto would be interesting: with the computer lyrics perhaps standing out as the more 'original', metaphorical, daring or unusual. To assist further, the Cambridge team also completed a statistical analysis of lyric word frequencies across different musical theatre song types designated by the team (Angry, I Am, I Want, Love, Comedy, Comfort, Duets, Protest) and gave the writers word clouds to access during curation and writing of the lyrics. The song *We are Greenham* was entirely constructed from lyric word clouds made from the real protest songs sung by the original Greenham Common peace protesters and another word cloud of lyrics used in protest songs in musical theatre.

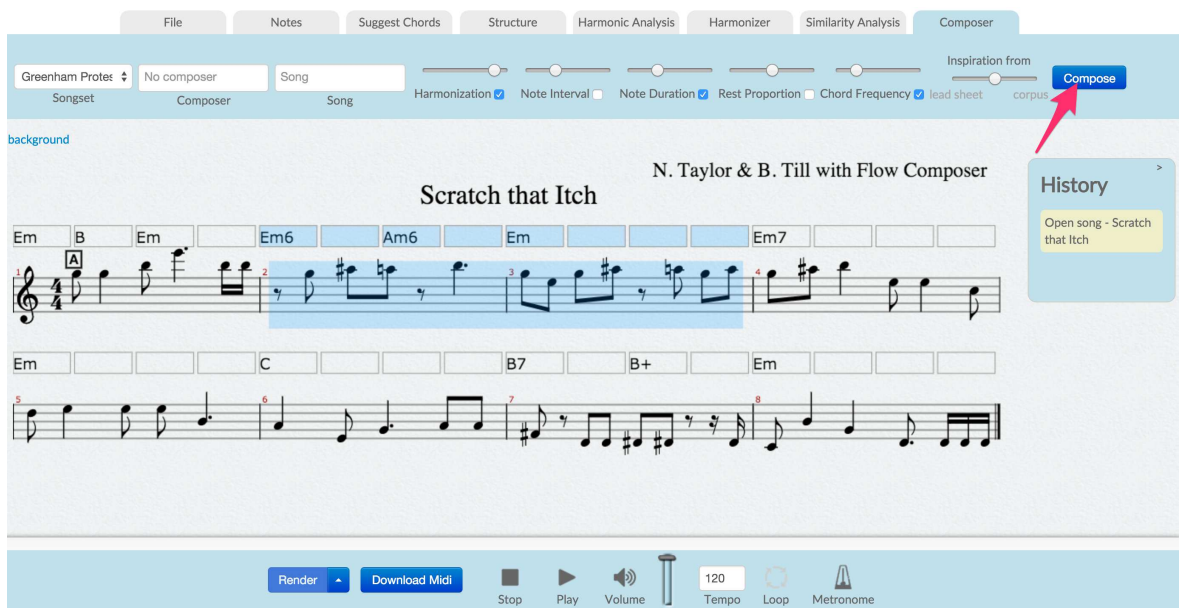

Figure 5: The web interface for Flow Composer. The Compose button generates music for the selected (highlighted) part.

## Impact

*Beyond the Fence* had 15 performances between 22nd February and March 3rd 2016. 3 performances were previews before the official opening, 2 were matinees, and 1 was a gala event. 3,047 people saw the musical, representing 59% of the total capacity across the run, which is impressive, given that there was almost no significant advertising or paid marketing, unlike for most musicals of this scale. For the last three performances, the audiences were polled and asked to rate their enjoyment of the show from 1 (low) to 5 (high). Of the 57 respondents, the poll revealed an overwhelmingly high level of enjoyment, with rank/percentages as follows: 1/1.7%, 2/1.7%, 3/10.3%, 4/17.3% and 5/69.0%.

Press engagement was managed in two phases: first with the announcement of the project/on-sale for tickets (December 2015); second in the run-up to the opening of the show and transmission of the documentaries (February 2016). *Beyond the Fence* and *Computer Says Show* received extensive domestic and international coverage: in print and online (including but not limited to The Guardian, The Times, The Evening Standard, The Independent, The Telegraph, BBC News Online, Financial Times, Daily Mail, New Scientist and Vice); and on television and radio (BBC Breakfast News, BBC Radio 4-World at One, CBC Radio-Canada, Sky News, Reuters TV, NHK-Japan, CBS News-US and the Guardian Tech Podcast). Interviews were conducted with the television and theatre production teams, the two writers, and some of the academics involved.

Reports from stage management document consistently engaged and reactive audiences, and reactions on social media have also been very positive. Audience members tend to express surprise that they are moved – often to tears – by parts of the show (with some expressing surprise that computers could have been involved in triggering their emotions, others insisting that the emotional sections must have been down to the human inputs, and some feeling 'more manipulated' than they would otherwise because they know computers were involved in playing with their emotions).

Thanks to footfall in the West End, and the frequency of walk-in trade, on several occasions people have attended the show not actually knowing about its computational genesis. On one such an occasion a gentleman remained for a post-show Q&A 'with the scientists', and raised his hand to tell everyone he "... had no idea about all of this, and I'm just amazed – I thought it was brilliant, so well done". Another audience member, on the same evening added, in relation to a discussion around the challenges that Till and Taylor faced "I thought it was brilliant, and if you've done this when it was all so hard, I can only imagine that the next one – when the computers get better – is going to be even better!" Various images depicting *Beyond the Fence* and *Computer Says Show* are given in figure 6.

The following is an extract from an interview conducted with the two writers of the musical, which appeared in the programme for the show:

**Writers' Perspective** (Benjamin Till and Nathan Taylor)
Collaborating with computers is utterly unlike anything either of us have encountered before, and at times, it has been incredibly frustrating. The systems for ideation, and for sparking the inspiration of creativity in the human mind are brilliantly helpful, and often lead to ideas that we would never have come up with on our own. It gets trickier, however, when you get into the realms of generating actual material. In a musical, nothing is left to chance – every note, every word, every idea, needs to support and inform not only the overall story, but also the characters' individual journeys, and a million other things that all need to feed into each other. With these computer systems, nothing is bespoke. We waded through probably a thousand pages of computer-generated tunes to find the fragments and phrases that felt right for the show's needs, and were suitable for developing into the songs that have become *Beyond the Fence*.

Once we had found the material we wanted to develop, it didn't make any difference that some of it had come from us, and some from the computers. It all went through the same process of refinement and evolution as any other songs we've ever written. As a result of that refinement and development, we both feel every bit as much ownership of the piece as if we had written it all ourselves. It would be hard not to, as we have had to immerse ourselves in it, and put as much of ourselves into it as with anything else.

*Beyond the Fence* really is a world's first, and a piece of history in the making – it is a true collaboration between humans and computers, bringing together human endeavour and the best that technology can currently offer in the field of Computational Creativity. These computer processes are still very much in their infancy, and we feel privileged to be the vanguard of this kind of work. In a few years, who knows: you might be able to push a button, and out pops My Fair Lady, but somehow, I doubt it. I rather think that the future holds ways of allowing human artists to work with computers more comfortably, and with more control of their output, ultimately to support and perhaps shape their own creativity in ways they might not have been able to envisage.

### Critical Reviews

The critics were divided in terms of how they evaluated *Beyond the Fence* – some as they would any new musical; some focusing on the computer-generated nature of the material. There seems to have been a desire to pick apart the content in the show, in order to evaluate the 'more human' and 'more computer'-generated parts separately. The following *Broadway Baby* review highlights this tendency well:

"Their solos (everyone has a 'tick-box' solo) are the absolute highlights – particularly Matthewson who manages to create pathos through the roller-skating based song Graceful and gets what seems to be the most genuine applause of the night with a performance that is resonant of Victoria Wood's play on word delivery with Julie Walter's comic timing. It's interesting to note that Graceful is the song that had the fewest lyrics delivered by the computers (only 6%)."

Interestingly, the song *Graceful* was in fact inspired by one initial computer-generated line ('Now let me be fat'), from which a full lyric was written. Music was then algorithmically composed to the lyrics, lending the song much of its unusual rhythm, developed by the writers to become a song that brings the house down every night. This is just one example, but it indicates the challenge for reviewers in formulating a response, when attending to the material with both human and computer 'agency' in mind. This was expressed in more general terms in The Independent:

"... I wonder if the computer-generated tag will help or hinder: it's hard to think you'd watch the show without being more interested in the process than the product. And am I being romantic in thinking it's telling that while the story and songs work fine, the thing that makes it zing is the human-chosen setting?"

The critical response has been extensive and represents an unprecedented volume of expert reaction to, and opinion on, work in this field. The following are a selection of quotes from some of the reviews *Beyond the Fence* received:

*** "A unique experiment in musical theatre composition"
*The Stage*

*** "Despite my reservations I was won over"
*The Independent*

** "Hokey, but effective" *The Times*

** "As a theatrical event the show is remarkable"
*West End Frame*

*** "Wins you over with the weight of its clichés"
*Time Out*

*** "What's our measure for success? Well, the audience was applauding just about every number and was brought to tears" *The Londonist*

**** "Extremely moving and emotional ... it could be one of the most important pieces of theatre to come out of London this year" *West End Wilma*

*** "It may not herald a brave new world, but it does work as a night out ... in a world where flops are the norm, no mean feat" *Daily Telegraph*

Unsurprisingly, the more harsh reviews focused on the formulaic nature of the musical imposed through the machine learning exercises. These included the following:

** "Guess what? When you get a computer to create a musical, as Sky Arts has done – using data from the structure, scores and scripts of hundreds of musicals to generate scenarios, melodies and lyrics – it sounds just like a musical composed by a computer. This show is as bland, inoffensive, and pleasant as a warm milky drink." *Guardian*

** "Isn't it obvious? If you take the average of a load of hit musicals, you'll end up with something pretty average. Follow a formula and – who would have thunk it – you get something formulaic." *What's On Stage*

## Conclusions

This paper acts primarily as a record of the project which led to the *Beyond the Fence* musical and *Computer Says Show* documentaries. This project stands as perhaps the largest ever application – in the sense of breadth of software employed and the scale of achievement – of generative software for cultural benefit, and an important experiment in the usage of Computational Creativity research prototypes by non-experts. While the number of systems employed and the scale of the output that the software influenced – an entire West End musical theatre production – were impressive, the software was sometimes a hindrance rather than a benefit to the writers. In most instances, the software acted as more of a muse and/or creativity support tool to the two writers of the musical rather than as the primary author. Computational Creativity exists as a research field to help bring about a future where creative software positively affecting our lives is as commonplace as the benefits of telecommunications systems or social networking technologies. This is not going to happen through scientific experimentation alone: the challenge of building and utilising creative software needs to be taken up by larger portions of society, including technology industries, creative industries and the arts.

Due to the recent completion of the project, it has not yet been possible to fully scientifically evaluate questions such as the level of creative contribution each piece of software made, how easy/difficult the writers found using software to guide their creative process rather than merely enabling it, or whether the musical was a popular and/or critical success. All these questions merit further research. However, from a cultural point of view, the project has clearly been successful. This can be measured in terms of: the increase in diversity of audiences for our research outputs, to include musical theatre lovers; the sheer number of people exposed to ideas from Computational Creativity research through the press and television news reviews and the Sky Arts documentaries; and the fact that there has, to the best of our knowledge, been no societal backlash against the idea of software contributing creatively to cultural projects. Such projects as the one described here help to bring about acceptance of the notion of software being our creative partners, adding to cultural life in useful and meaningful ways, and ultimately bringing many benefits across society.

## Acknowledgements

The contribution of The WhatIf Machine to this project has been supported by EC grant WHIM (FP7 grant 611560), and an EPSRC Leadership Fellowship grant (EP/J004049). We would like to thank Catherine Bellamy for excellent work in supporting the link between WHIM project members and the Wingspan Productions team. We would also like to thank the organisers of the PROSECCO network (FP7 grant 600653) and particularly the organisers of the 2015 *Show, Tell, Imagine* event at Queen Mary University of London, where the writers of *Beyond the Fence* and the Wingspan Productions team met to discuss potential collaboration. Flow Machines is funded by the European Research Council (ERC) under the European Union 7th Framework Programme (FP7/2007-2013), ERC Grant Agreement number 291156. We thank the anonymous reviewers for their very helpful comments.

*Computer Says Show* (Gale, Baron, and Lomax 2016) and *Beyond the Fence* (Till, Taylor et al. 2016) were commissioned by Sky Arts, who were the majority funders (Director Phil Edgar-Jones and Commissioning Editor Siobhan Mulholland). It was developed with the assistance of the Wellcome Trust. It was produced by Wingspan Productions. The Head of Production for the project was Lil Cranfield. *Beyond the Fence* was first performed at the Arts Theatre London on 22nd February 2016. The General Manager was Neil Laidlaw Productions.

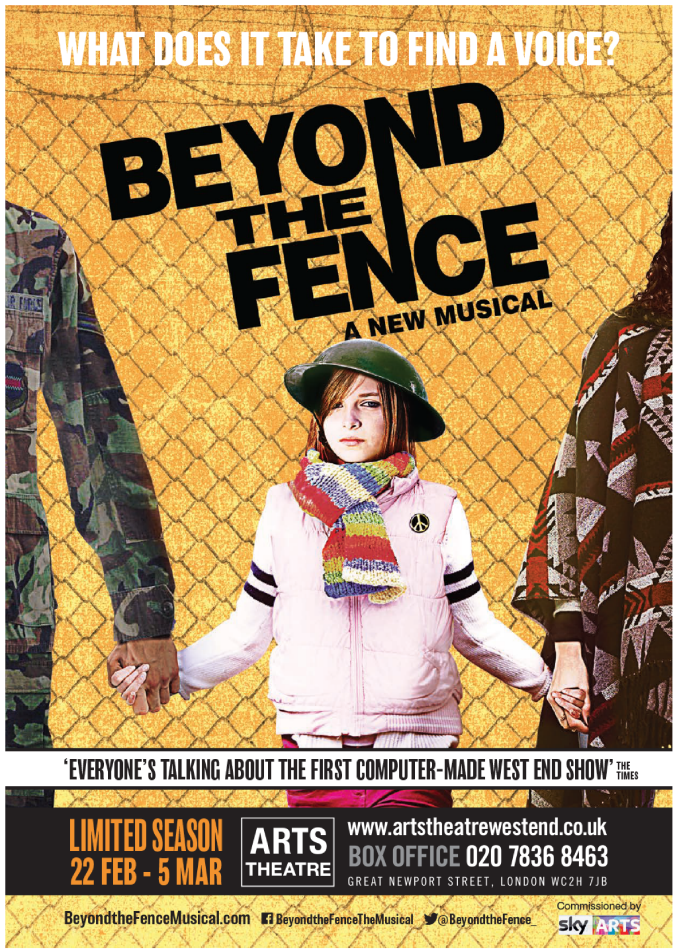
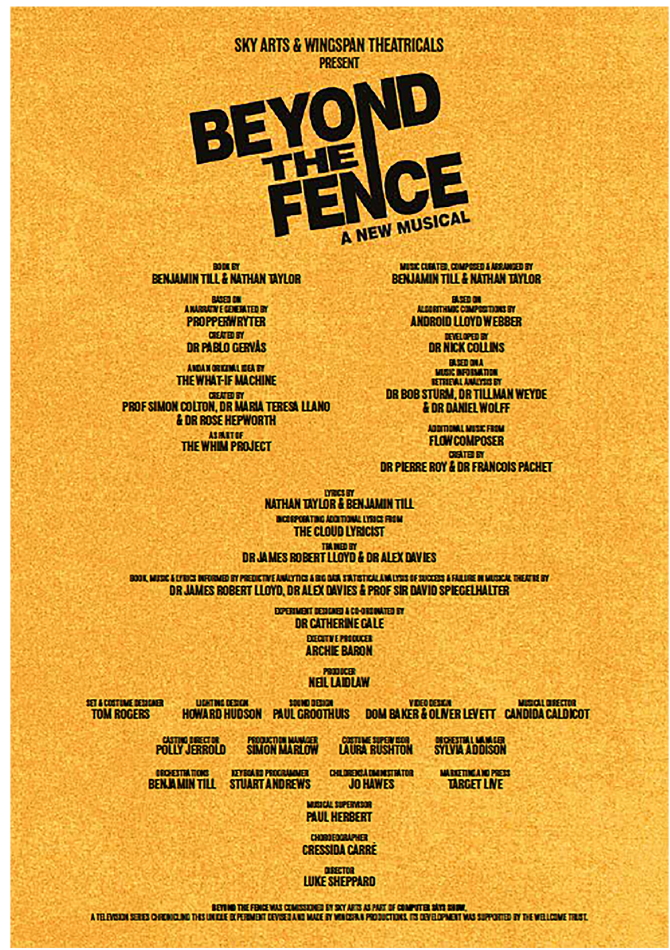
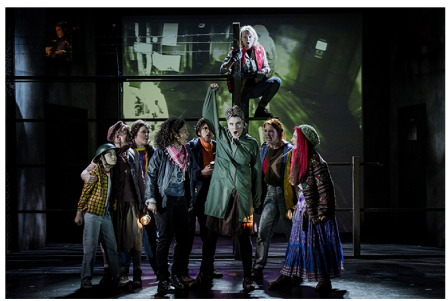
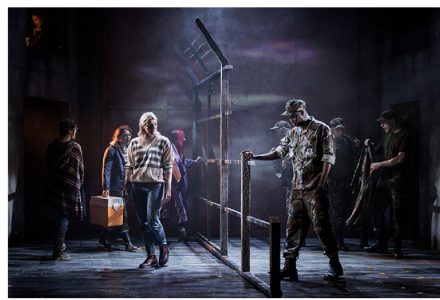
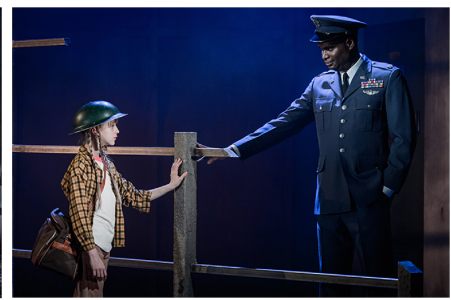
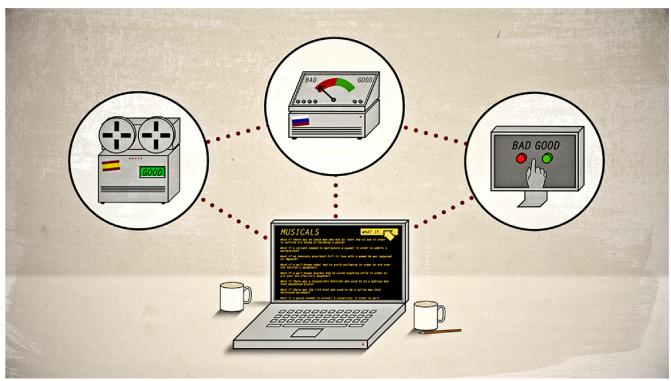
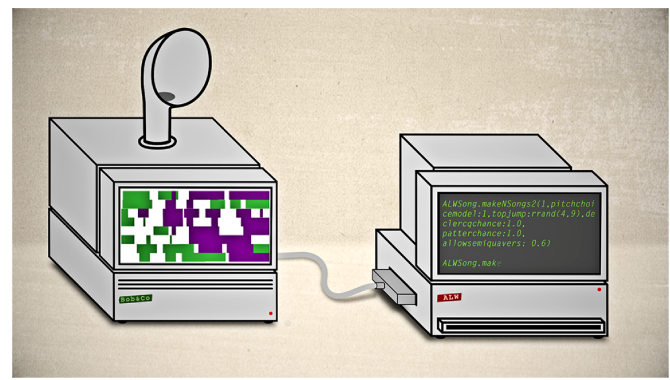

Figure 6: (a) The poster for the show (b) the creative billing, where the software systems take their rightful place among the team that developed the musical (c) images from the musical *Beyond the Fence* (credit: Robert Workman), and (d) graphics used in the *Computer Says Show* documentary (credit: Andy A'Court).